\documentclass[preprint,showpacs,preprintnumbers,amsmath,amssymb]{revtex4}
\usepackage{amssymb}
\usepackage{txfonts}
\usepackage{graphicx}
\usepackage{dcolumn}
\usepackage{bm}
\usepackage{mathrsfs}

\begin{document}

JHEP 0808 (2008) 041

\title{Localization and Mass Spectrum of Matters on Weyl Thick Branes}

\author{Yu-Xiao Liu\footnote{Corresponding author. Email address: liuyx@lzu.edu.cn},
        Li-Da Zhang, Shao-Wen Wei and Yi-Shi Duan}

\affiliation{
 Institute of Theoretical Physics, Lanzhou University,
 Lanzhou 730000, P. R. China}

\begin{abstract}
In this paper, we study localization and mass spectrum of various
matter fields on a family of thick brane configurations in a pure
geometric Weyl integrable 5-dimensional space time, a non-Riemannian
modification of 5-dimensional Kaluza--Klein (KK) theory. We present
the shape of the mass-independent potential of the corresponding
Schr\"{o}dinger problem and obtain the KK modes and mass spectrum,
where a special coupling of spinors and scalars is considered for
fermions. It is shown that, for a class of brane configurations,
there exists a continuum gapless spectrum of KK modes with any
$m^2>0$ for scalars, vectors and ones of left chiral and right
chiral fermions. All of the corresponding massless modes are found
to be normalizable on the branes. However, for a special of brane
configuration, the corresponding effective Schr\"{o}dinger equations
have modified P\"{o}schl-Teller potentials. These potentials suggest
that there exist mass gap and a series of continuous spectrum
starting at positive $m^2$. There are one bound state for spin one
vectors, which is just the normalizable vector zero mode, and two
bound KK modes for scalars. The total number of bound states for
spin half fermions is determined by the coupling constant $\eta$. In
the case of no coupling ($\eta=0$), there are no any localized
fermion KK modes including zero modes for both left and right chiral
fermions. For positive (negative) coupling constant, the number of
bound states of right chiral fermions is one less (more) than that
of left chiral fermions. In both cases ($\eta>0$ and $\eta<0$), only
one of the zero modes for left chiral fermions and right chiral
fermions is bound and normalizable.
\end{abstract}

\pacs{11.10.Kk., 04.50.+h.  \\
Keywords: Large Extra Dimensions,
          Field Theories in Higher Dimensions}


\maketitle

\section{Introduction}

Suggestions that extra dimensions may not be compact
\cite{rs,Lykken,RubakovPLB1983136,RubakovPLB1983139,Akama1983,VisserPLB1985,Randjbar-DaemiPLB1986}
or large \cite{AntoniadisPLB1990,ADD} can provide new insights for a
solution of gauge hierarchy problem \cite{ADD} and cosmological
constant problem
\cite{RubakovPLB1983139,Randjbar-DaemiPLB1986,KehagiasPLB2004}. In
the framework of brane scenarios, gravity is free to propagate in
all dimensions, whereas all the matter fields are confined to a
3--brane
\cite{ADD,RubakovPLB1983139,VisserPLB1985,SquiresPLB1986,gog}. In
Ref. \cite{rs}, an alternative scenario of the compactification has
been put forward. In the brane world scenario, an important question
is localization of various bulk fields on a brane by a natural
mechanism. It is well known that the massless scalar field
\cite{BajcPLB2000} and the graviton \cite{rs} can be localized on
branes of different types, and that the spin 1 Abelian vector fields
can not be localized on the Randall-Sundrum(RS) brane in five
dimensions but can be localized on some branes in higher dimensions
\cite{OdaPLB2000113}. For spin 1/2 fermions, they do not have
normalizable zero modes in five and six dimensions
\cite{BajcPLB2000,OdaPLB2000113,NonLocalizedFermion,Ringeval,KoleyCQG2005,GherghettaPRL2000,Neupane}.

Recently, an increasing interest has focused on study of thick
brane scenarios based on gravity coupled to scalars in higher
dimensional space-time
\cite{dewolfe,gremm,Csaki,varios,ThickBrane_Dzhunushaliev}. An
interesting feature of these models is that one can obtain branes
naturally without introducing them by hand in the action of the
theory \cite{dewolfe}. Furthermore, these scalar fields provide
the ``material" from which the thick branes are made of. Thirdly,
for the branes with inclusion of scalar backgrounds
\cite{RandjbarPLB2000}, localized chiral fermions can be obtained
under some conditions.

In this paper, we are interested in the thick branes based on
gravity coupled to scalars in a Weyl integrable manifold
\cite{ThickBrane1,ThickBrane2,ThickBrane3,Liu0708,ThickBrane4}. In
this scenario, spacetime structures with pure geometric thick smooth
branes separated in the extra dimension arise. For most of these
branes, there exists a single bound state which represents a stable
4-dimensional graviton and the spectrum of massive modes of
Kaluza--Klein (KK) excitations is continuous without mass gap
\cite{ThickBrane2,ThickBrane3}. This gives an very important
conclusion: the claim that Weylian structures mimic classically
quantum behavior does not constitute a generic feature of these
geometric manifolds \cite{ThickBrane2}. In Ref. \cite{ThickBrane4},
it is shown that, for one of these branes, there exist one massless
bound state (the massless 4-dimensional grivaton), one massive KK
bound state and the continuum spectrum of delocalized KK modes. The
mass hierarchy problem and the corrections to Newton's law in the
thin brane limit was considered.

In our previous work \cite{Liu0708}, we studied localization of
various matter fields on some of these pure geometrical Weyl thick
branes. It is shown that, for both scalars and vectors, there exists
a single bound state and a continuum gapless spectrum of massive KK
states. But only the massless mode of scalars is found to be
normalizable on the brane. For the massless fermions localization,
there must have some kind of Yukawa coupling. The aim of the present
article is to investigate localization of various matters on one of
the pure geometrical thick branes obtained in Refs.
\cite{ThickBrane1,ThickBrane2,ThickBrane3}. We will show that, in
the brane model, there exist one and two discrete bound states for
scalars and vectors respectively (the ground states are normalizable
massless modes), and a series of continuum massive KK states for
both fields. For spin 1/2 fermions, there are a finite number of
bound states which depend on the coupling constant and only one of
the zero modes for left and right chiral fermions is bound. The
paper is organized as follows: In section \ref{SecModel}, we first
give a review of the thick brane arising from a pure geometric Weyl
integrable 5-dimensional space time, which is a non-Riemannian
modification of 5-dimensional KK theory. Then, in section
\ref{SecLocalize}, we study localization of various matter fields
with spin ranging from 0 to 1 on the pure geometrical thick brane in
5 dimensions by presenting the shape of the potentials of the
corresponding Schr\"{o}inger equations. Finally, a brief conclusion
and discussion are presented.

\section{Review of Weyl thick branes}
\label{SecModel}


Let us start with a pure geometrical Weyl action in five
dimensions---a non--Riemannian generalization of KK theory
\begin{equation}
\label{action} S_5^W =\int_{M_5^W}\frac{d^5x\sqrt{-g}} {16\pi G_5}
e^{\frac{3}{2}\omega} \left[R+3\tilde{\xi}(\nabla\omega)^2 +
6U(\omega)\right],
\end{equation}
where $M_5^W$ is a 5-dimensional Weyl-integrable manifold
specified by the pair $(g_{MN},\omega)$, and $\omega$ is a Weyl
scalar function. The Weylian Ricci tensor is given by
$R_{MN}=\Gamma_{MN,P}^P-\Gamma_{PM,N}^P
+\Gamma_{MN}^P\Gamma_{PQ}^Q-\Gamma_{MQ}^P\Gamma_{NP}^Q$, with
$\Gamma_{MN}^P = \{_{MN}^{\;\;P}\}  - \frac{1}{2} (\omega_{,M}
\delta_N^P+\omega_{,N} \delta_M^P-g_{MN}\omega^{,P})$ the affine
connections on $M_5^W$ and $\{_{MN}^{\;\;P}\}$ the Christoffel
symbols. The parameter $\tilde{\xi}$ is a coupling constant, and
$U(\omega)$ is a self-interaction potential for $\omega$. The Weyl
action is of pure geometrical nature since the scalar field
$\omega$ enters in the definition of the affine connections of the
Weyl manifold. The line-element which results in a 4-dimensional
Poincar$\acute{e}$ invariance of the Weyl action (\ref{action}) is
assumed as follows
\begin{equation}
\label{linee} ds_5^2=e^{2A(y)}\eta_{\mu\nu}dx^\mu dx^\nu + dy^2,
\end{equation}
where $e^{2A(y)}$ is the warp factor, and $y$ is the extra
coordinate. Solutions can be found in Refs.
\cite{ThickBrane1,ThickBrane2,ThickBrane3,Liu0708}. Here, we only
reconsider the following one which corresponds to the
self-interaction potential $U=\lambda e^{p\;\omega}$
\begin{eqnarray}
 e^{2A(y)}   &=& \left[\cos\left(\sqrt{8{\lambda}p}(y-c)\right)\right]^{\frac{3}{2p}},\\
 e^{\omega}~~ &=& \left[\cos\left(\sqrt{8{\lambda}p}(y-c)\right)\right]^{-\frac{2}{p}}.
  \label{configuration1}
\end{eqnarray}
where $p=16\tilde{\xi}-15$, $\lambda$ and $c$ are arbitrary
constants. The solution results in a compact manifold along the
extra dimension with range
$-\frac{\pi}{2}\leq\sqrt{8{\lambda}p}(y-c)\leq\frac{\pi}{2}$, and
describes a single Weyl thick brane. For more details, one can
refer to Refs.
\cite{ThickBrane1,ThickBrane2,ThickBrane3,ThickBrane4}.


\section{Localization of various matters}
\label{SecLocalize} Since gravitons can be localized on the Weyl
thick brane described in previous section, it is a pertinent
question to ask whether various bulk mater fields such as scalars,
spin one vector fields and spin 1/2 fermions can be localized on the
Weyl thick brane by means of only the gravitational interaction. We
will analyze the spectrums of various mater fields for the thick
brane by present the potential of the corresponding Schr\"{o}dinger
equation. In order to get mass-independent potential, we will change
the metric given in (\ref{linee}) to a conformally flat one
\begin{equation}
\label{conflinee2} ds_5^2=e^{2A}\left(\eta_{\mu\nu}dx^\mu
dx^\nu+dz^2\right)
\end{equation}
by performing the coordinate transformation
\begin{equation}
dz=e^{-A(y)}dy. \label{transformation}
\end{equation}

\subsection{Spin 0 scalar field}

Let us first consider the action of a massless real scalar coupled
to gravity
\begin{eqnarray}
S_0 = - \frac{1}{2} \int d^5 x  \sqrt{-g}\; g^{M N}
\partial_M \Phi \partial_N \Phi.
\label{scalarAction}
\end{eqnarray}
By considering the conformally flat metric  (\ref{conflinee2}) the
equation of motion which can be derived from (\ref{scalarAction})
is read
\begin{eqnarray}
\left( \partial^2_z + 3(\partial_{z} A) \partial_z
       +\eta^{\mu\nu} \partial_\mu \partial_\nu
 \right) \Phi = 0. \label{scalarEOM}
\end{eqnarray}
Then, by decomposing $\Phi(x,z) =   \sum_n \phi_n(x)\chi_n(z)$ and
demanding $\phi_n(x)$ satisfies the 4-dimensional massive
Klein--Gordon equation $(\eta^{\mu\nu}\partial_\mu \partial_\nu
-m_n^2 )\phi_n(x)=0 $, we obtain the equation for $\chi_n(z)$
\begin{eqnarray}
\left( \partial^2_z + 3 (\partial_{z} A) \partial_z
       +m_n^2 \right) \chi_n(z) = 0.
 \label{massiveScalar1}
\end{eqnarray}
The full 5-dimensional action (\ref{scalarAction}) reduces to the
standard 4-dimensional action for the massive scalars, when
integrated over the extra dimension under the conditions that the
above equation is satisfied and the following normalization
condition is obeyed
\begin{eqnarray}
 \int^{\infty}_{-\infty} dz
 \;e^{3A}\chi_m(z)\chi_n(z)=\delta_{mn}.
 \label{normalizationCondition1}
\end{eqnarray}

\begin{figure}[htb]
\begin{center}
\includegraphics[width=7.5cm,height=5.5cm]{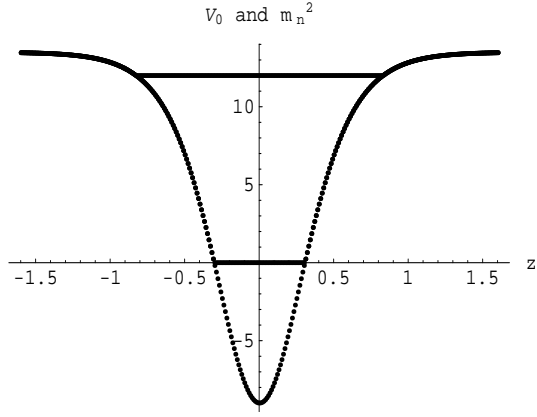}
\end{center}
\caption{The shape of the potentials $V_0$ and the mass spectrum
$m_n^2$ for scalars for the case $p=3/4$. The parameter is set to
$\lambda=1$.}
 \label{fig_Vz_Mn2_Scalar}
\end{figure}

By defining $\widetilde{\chi}_n(z)=e^{\frac{3}{2}A}\chi_n(z)$, the
wave function (\ref{scalarEOM}) can be recast in the form of a
Schr\"{o}dinger equation
\begin{eqnarray}
  \left[-\partial^2_z+ V_0(z)\right]\widetilde{\chi}_n(z)=m_n^2 \widetilde{\chi}_n(z),
  \label{SchEqScalar1}
\end{eqnarray}
where $m_n$ is the mass of the KK excitation and the effective
potential is given by
\begin{eqnarray}
  V_0(z)=\frac{3}{2}\partial^2_z A + \frac{9}{4}(\partial_z A)^2.
\end{eqnarray}
The potential has the same form as the case of graviton and has been
discussed detailedly in \cite{ThickBrane4}. Here we only give the
main results. In order to map the compact $y$-interval onto the real
$z$-line, we require that $0<p\leq 3/4$, which results in that
$z(y)$ is a monotonous function and $z\rightarrow \pm \infty$ when
$\sqrt{8{\lambda}p}(y-c)\rightarrow \pm\frac{\pi}{2}$. For
$0<p<3/4$, the limit of the potential $V_0(z)$ is zero when
$z\rightarrow \pm \infty$, which results in that the Schr\"{o}dinger
equation (\ref{SchEqScalar1}) will have a continuous spectrum
starting at zero and only the massless mode is bound. For the case
$p=3/4$, the limit of the potential $V_0(z)$ is the finite value
$V_0(\infty)=27\lambda/2$, and one can invert the coordinate
transformation $dz=e^{-A(y)}dy$: $\cos[\sqrt{6\lambda}(y-c)]={\rm
sech}\bigl[\sqrt{6\lambda}(z-z_0)\bigr]$. The potential becomes a
modified P\"{o}schl-Teller potential
\begin{eqnarray}
V_0(z) =\frac{9\lambda}{2} \Bigl\lbrace 3 - 5\, {\rm sech}^2\Bigl[
\sqrt{6\lambda}(z-z_0)\Bigr] \Bigr\rbrace.
\end{eqnarray}
For the potential, there are two bound states (see, e.g.,
\cite{poshltellerp}). One is the normalizable ground state
\begin{eqnarray}
\widetilde{\chi}_0(z)=c_0 {\rm sech}^{3/2}(\sqrt{6\lambda}(z-z_0))
\end{eqnarray}
with mass $m_0^2=0$ (zero-mass state), another is the normalizable
exited state
\begin{eqnarray}
\widetilde{\chi}_1(z)=c_1 \sinh (z){\rm
sech}^{3/2}(\sqrt{6\lambda}(z-z_0))
\end{eqnarray}
with mass $m_1^2=12\lambda$. The continuous spectrum start with $m^2
= \frac{27}{2} \lambda$ and asymptotically turn into plane waves,
which represent delocalized KK massive scalars. The potential $V_0$
and the mass spectrum are showed in Fig. \ref{fig_Vz_Mn2_Scalar}.

\subsection{Spin 1 vector field}

Next we turn to spin 1 vector fields. Here we consider the action
of $U(1)$ vector fields
\begin{eqnarray}
S_1 = - \frac{1}{4} \int d^5 x \sqrt{-g} g^{M N} g^{R S} F_{MR}
F_{NS}, \label{actionVector}
\end{eqnarray}
where $F_{MN} = \partial_M A_N - \partial_N A_M$ as usual. From
this action and the background geometry (\ref{conflinee2}), the
equation motion is given by
\begin{eqnarray}
 \eta^{\mu\nu} \partial_\mu F_{\nu 4} &=& 0, \\
 \partial^\mu F_{\mu \nu}
      + \left(\partial_z + \partial_{z} A \right) F_{4 \nu}&=& 0.
\label{46}
\end{eqnarray}
We assume that the $A_\mu$ are $Z_2$-even and that $A_4$ is
$Z_2$-odd with respect to the extra dimension $z$, which results
in that $A_4$ has no zero mode in the effective 4D theory.
Furthermore, in order to consistent with the gauge invariant
equation $\oint dz A_4=0$, we use gauge freedom to choose $A_4=0$.
Under these assumption, the action (\ref{actionVector}) is reduced
to
\begin{eqnarray}
S_1 = - \frac{1}{4} \int d^4 x dz
  \left(e^A \eta^{\mu\lambda} \eta^{\nu\rho}
                 F_{\mu\nu} F_{\lambda\rho}
        -2 \eta^{\mu\nu}A_{\mu}\partial_z
        \left( e^A \partial_z A_{\nu}  \right)
  \right).
\label{actionVector2}
\end{eqnarray}
Then, with the decomposition of the vector field
$A_{\mu}(x,z)=\sum_n a^{(n)}_\mu(x)\rho_n(z)$, and importing the
normalization condition
\begin{eqnarray}
 \int^{\infty}_{-\infty} dz \;e^A\rho_m(z)\rho_n(z)=\delta_{mn},
 \label{normalizationCondition2}
\end{eqnarray}
the action (\ref{actionVector2}) is read
\begin{eqnarray}
S_1 = \sum_n \int d^4 x
  \left( - \frac{1}{4} \eta^{\mu\lambda} \eta^{\nu\rho}
                 f^{(n)}_{\mu\nu} f^{(n)}_{\lambda\rho}
        -\frac{1}{2}m_n^2 \eta^{\mu\nu}a^{(n)}_{\mu}a^{(n)}_{\nu}
  \right),
\label{actionVector3}
\end{eqnarray}
where $f^{(n)}_{\mu\nu} = \partial_\mu a^{(n)}_\nu - \partial_\nu
a^{(n)}_\mu$ is the 4-dimensional field strength tensor, and it
has been required that the $\rho_n(z)$ satisfies the equation
\begin{eqnarray}
(\partial^2_z +(\partial_{z} A) \partial_z+m_n^2) \rho_n(z) =0.
\label{diffEqVector}
\end{eqnarray}

\begin{figure}[htb]
\begin{center}
\includegraphics[width=7.5cm,height=5.5cm]{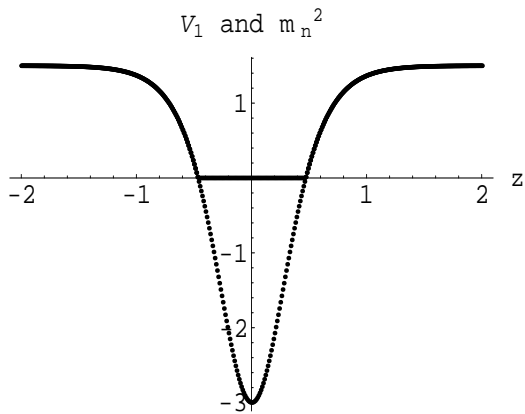}
\end{center}
\caption{The shape of the potentials $V_1$ and the mass spectrum
$m_n^2$ for vectors for the case $p=3/4$. The parameter is set to
$\lambda=1$.}
 \label{fig_Vz_Mn2_Vector}
\end{figure}

By defining $\widetilde{\rho}_n=e^{A/2}{\rho}_n$, we get the
corresponding Schr\"{o}dinger equation for the vector field
\begin{eqnarray}
  \left[-\partial^2_z +V_1(z) \right]\widetilde{\rho}_n(z)=m_n^2
  \widetilde{\rho}_n(z),  \label{diffEqVector2}
\end{eqnarray}
where the potential is given by
\begin{eqnarray}
  V_1(z)= \frac{1}{2}\;\partial^2_z A
       +\frac{1}{4}\;(\partial_z A)^2 .
\end{eqnarray}
The potential can be calculated as a function of $y$ by using the
coordinate transformation (\ref{transformation}).
\begin{eqnarray}
  V_1(z(y))= \frac{3 \lambda}{8 p}
   \cos^{3/2p-2} ( \sqrt{8\lambda p}\; y)
   \left(9\sin^2(\sqrt{8\lambda p}\; y)-8p\right),
\end{eqnarray}
where we have set $c=0$. For $0<p\leq3/4$, $z(y)$ is a monotonous
function, which implies that
\begin{eqnarray}
\lim_{z\to \pm \infty} V_1(z)=\lim_{\sqrt{8\lambda p}\;y\to \pm
{\frac{\pi}{2}}} V_1(z(y)).
\end{eqnarray}
This limit is zero for $0<p<3/4$ and $3\lambda/2$ for $p=3/4$. So
for the case $0<p<3/4$, the Schr\"{o}dinger equation
(\ref{diffEqVector2}) will have a continuous spectrum starting at
zero and only the massless mode is bound. For the case $p=3/4$,
one can invert the coordinate transformation $dz=e^{-A(y)}dy$ and
explicit form of the potential is turned out to be
\begin{eqnarray}
 V_1(z) =\frac{3}{2}\lambda \; \left(1
      -3 {\rm sech}^2(\sqrt{6\lambda}\;z) \right).
\end{eqnarray}
This potential has a minimum (negative value) $V_1(z=0)=-3 \lambda$
at the location of brane and the asymptotic behavior: $V_1(z=\pm
\infty)=3\lambda/2$, which implies that there is a mass gap. By
rescaling $u=\sqrt{6\lambda}\;z$, Eq. (\ref{diffEqVector2}) turns
into the well-known Schr\"{o}dinger equation with $\nu=1/2$ and
$E_n=m_n^2/6\lambda-1/4$
\begin{eqnarray}
\Bigl[-\partial_u^2  - \nu(\nu+1){\rm sech}^2(u)\Bigr]~
\widetilde{\rho}_n = E_n ~ \widetilde{\rho}_n .
\end{eqnarray}
For this equation with a modified P\"{o}schl-Teller potential,
there is only one bound state, i.e., the ground state
\begin{eqnarray}
\widetilde{\rho}_0(z)=\frac{(6\lambda)^{1/4}}{\sqrt{\pi}} {\rm
sech}^{1/2}(\sqrt{6\lambda}\;z)
\end{eqnarray}
with energy $E_0=-\nu^2=-1/4$, which is just the normalized
zero-mass mode and also shows that there is no tachyonic vector
modes. The potential $V_1$ and the mass spectrum are showed in Fig.
\ref{fig_Vz_Mn2_Vector}. The continuous spectrum start with $m^2 =
\frac{3}{2} \lambda$ and asymptotically turn into plane waves, which
represent delocalized KK massive vectors.

It was shown in the RS model in $AdS_5$ space that a spin 1 vector
field is not localized neither on a brane with positive tension nor
on a brane with negative tension so the Dvali-Shifman mechanism
\cite{DvaliPLB1997} must be considered for the vector field
localization \cite{BajcPLB2000}. Here, it is turned out that a
vector field can be localized on the thick brane for $0<p\leq3/4$
and we do not need to introduce additional mechanism for the vector
field localization in the case at hand. For both cases of $0<p<3/4$
and $p=3/4$, there is only one bound state which is the zero-mass
mode. While for the latter case $p=3/4$, there exist a mass gap
between the ground state and the first exited state.

\subsection{Spin 1/2 fermionic field}\label{SecFermionic}

In this subsection, we will investigate whether spin half fermions
can be localized on the brane. In five dimensions, fermions are four
component spinors and their Dirac structure is described by
$\Gamma^M= e^M _{\bar{M}} \Gamma^{\bar{M}}$ with
$\{\Gamma^M,\Gamma^N\}=g^{MN}$. In our set-up,
$\Gamma^M=(e^{-A}\gamma^{\mu},e^{-A}\gamma^5)$, where $\gamma^{\mu}$
and $\gamma^5$ are the usual flat gamma matrices in the Dirac
representation. The Dirac action of a massless spin 1/2 fermion
coupled to gravity and scalar is
\begin{eqnarray}
S_{1/2} = \int d^5 x \sqrt{-g} \left(\bar{\Psi}  \Gamma^M D_M
\Psi-\eta \bar{\Psi} F(\omega) \Psi\right), \label{DiracAction}
\end{eqnarray}
where the covariant derivative $D_M$ is defined as $D_M\Psi =
(\partial_M + \frac{1}{4} \omega_M^{\bar{M} \bar{N}}
\Gamma_{\bar{M}} \Gamma_{\bar{N}} ) \Psi$ with the spin connection
$\omega_M= \frac{1}{4} \omega_M^{\bar{M} \bar{N}} \Gamma_{\bar{M}}
\Gamma_{\bar{N}}$. In this paper, $\bar{M}, \bar{N}, \cdots$
denote the local Lorentz indices, and $\Gamma^{\bar{M}}$ are the
flat gamma matrices in five dimensions. In background
(\ref{conflinee2}), the non-vanishing components of the spin
connection $\omega_M$ are
\begin{eqnarray}
  \omega_\mu = \frac{1}{2}(\partial_{z}A) \gamma_\mu \gamma_5. \label{eq4}
\end{eqnarray}
From the Dirac action and the above equation, the equation of
motion is given by
\begin{eqnarray}
 \left\{ \gamma^{\mu}\partial_{\mu}
         + \gamma^5 \left(\partial_z  +2 \partial_{z} A \right)
         -\eta\; e^A F(\omega)
 \right \} \Psi =0, \label{DiracEq1}
\end{eqnarray}
where $\gamma^{\mu} \partial_{\mu}$ is the Dirac operator on the
brane.

We are now ready to study the above Dirac equation for 5-dimensional
fluctuations, and write it in terms of 4-dimensional effective
fields. Because of the Dirac structure of the fifth gamma matrix
$\Gamma^5=\gamma^5$, we expect that left- and right-handed
projections of the four dimensional part to behave differently.
Thus, from the equation of motion (\ref{DiracEq1}), we will search
for the solutions of the general chiral decomposition
\begin{equation}
 \Psi(x,z) = \sum_n\psi_{Ln}(x) \alpha_{Ln}(z)+\sum_n\psi_{Rn}(x) \alpha_{Rn}(z),
\end{equation}
where $\psi_{Ln}(x)$ and $\psi_{Rn}(x)$ are the left-handed and
right-handed components of a 4-dimensional Dirac field, they are a
fixed basis and the $\psi_{Ln}$ and $\psi_{Rn}$ are dynamical, and
the sum over $n$ can be both discrete and continuous. To obtain the
defining equations for the basis functions $\psi_{Ln}(x)$ and
$\psi_{Rn}(x)$, we assume that $\psi_{L}(x)$ and $\psi_{R}(x)$
satisfy the 4-dimensional massive Dirac equations
$\gamma^{\mu}\partial_{\mu}\psi_{Ln}(x)=m_n\psi_{R_n}(x)$ and
$\gamma^{\mu}\partial_{\mu}\psi_{Rn}(x)=m_n\psi_{L_n}(x)$. Then
$\alpha_{Ln}(z)$ and $\alpha_{Rn}(z)$ satisfy the following coupled
eigenvalue equations
\begin{subequations}
\begin{eqnarray}
 \left \{ \partial_z+2\partial_{z}A
                  + \eta\;e^A F(\omega) \right \} \alpha_{Ln}(z)
  =  ~~m_n \alpha_{Rn}(z), \label{CoupleEq1a}  \\
 \left \{ \partial_z+2\partial_{z}A
                  - \eta\;e^A F(\omega) \right\} \alpha_{Rn}(z)
  =  -m_n \alpha_{Ln}(z). \label{CoupleEq1b}
\end{eqnarray}\label{CoupleEq1}
\end{subequations}
In order to obtain the standard four dimensional action for the
massive chiral fermions, we need the following orthonormality
conditions
\begin{eqnarray}
 \int_{-\infty}^{\infty} e^{4A}  \alpha_{Lm} \alpha_{Rn}dz
   &=& \delta_{LR}\delta_{mn}.
\end{eqnarray}
for $\alpha_{L_{n}}$ and $\alpha_{R_{n}}$.

\begin{figure}[htb]
\begin{center}
\includegraphics[width=7.5cm,height=5.5cm]{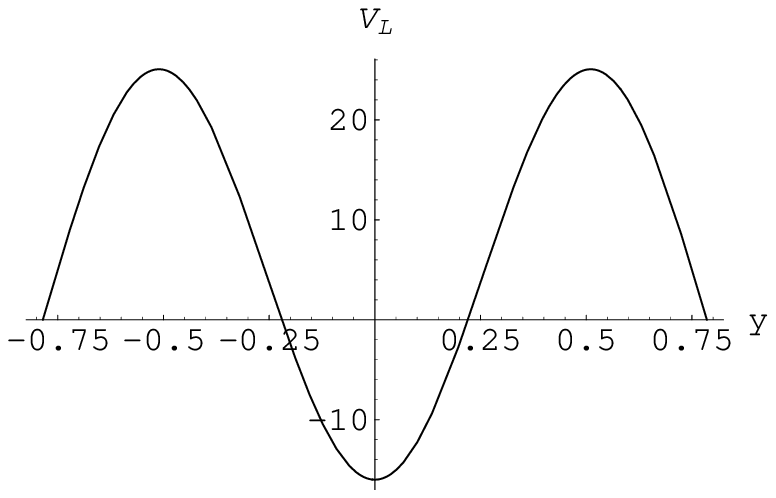}
\includegraphics[width=7.5cm,height=5.5cm]{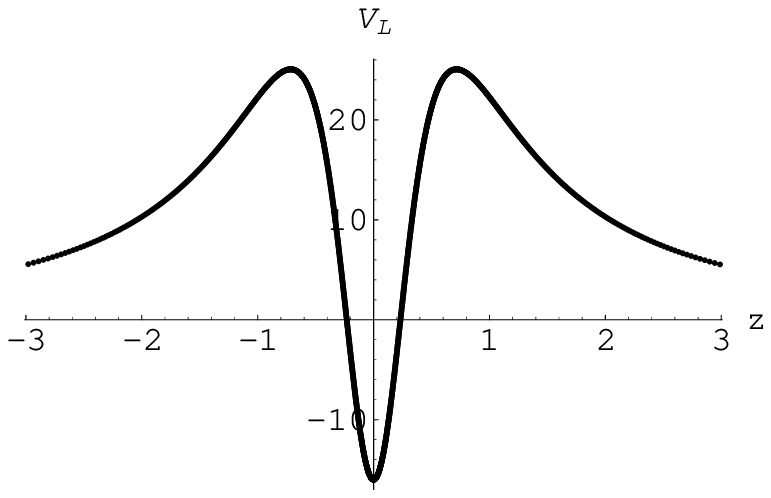}
\end{center}
\begin{center}
\includegraphics[width=7.5cm,height=5.5cm]{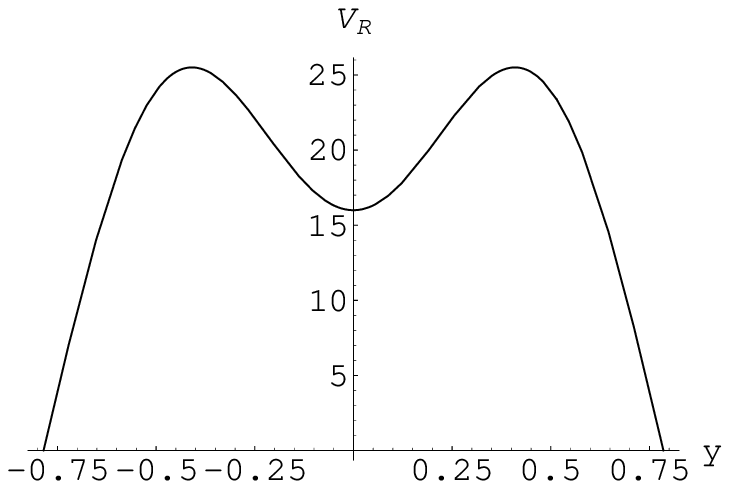}
\includegraphics[width=7.5cm,height=5.5cm]{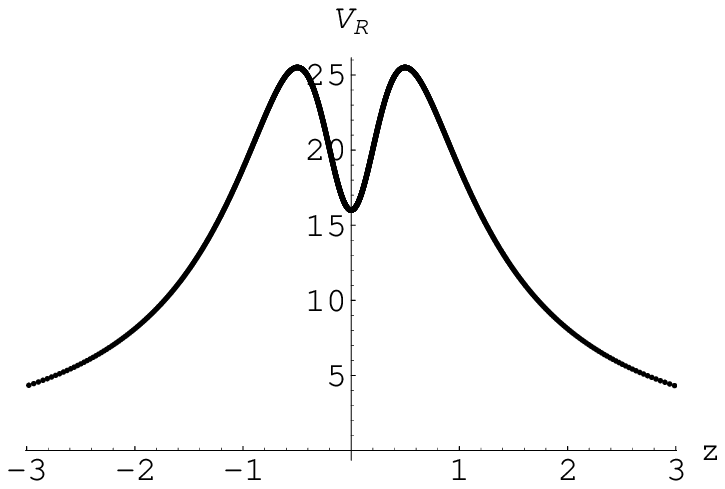}
\end{center}
\caption{The shape of the potentials $V_L$ and $V_R$ for left and
right chiral fermions for the case $0<p<3/4$ in $y$ and $z$
coordinates. The parameters are set to $p=1/2$, $\eta=1$ and
$\lambda=1$.}
 \label{fig_VzVy_Fermion}
\end{figure}

By defining $\widetilde{\alpha}_{Ln}=e^{2A}\alpha_{Ln}$, we get the
Schr\"{o}dinger equation for left chiral fermions
\begin{eqnarray}
  [-\partial^2_z + V_L(z) ]\widetilde{\alpha}_{Ln}=m_n^2 \widetilde{\alpha}_{Ln}
  \label{SchEqLeftFermion3}
\end{eqnarray}
with the effective potential
\begin{eqnarray}
  V_L(z)= e^{2A} \eta^2 F^2(\omega)
     - e^{A} \eta\; \partial_z F(\omega)
     - (\partial_{z}A) e^{A} \eta F(\omega).
\end{eqnarray}
For localization of left chiral fermions around the brane, the
effective potential $V_L(z)$ should have a minimum at the brane.
Furthermore, we also demand a symmetry for $V_L(z)$ about the
position of the brane. This requires $F(\omega(z))$ to be an odd
function of $z$. Since $\omega(z)$ is a even function, we set
$F(\omega(z))=e^{\frac{1}{2}p\;\omega(z)}\partial_{z}\omega(z)$ as
an example. Here, we face the difficulty again that for general
$p$ we can not solve the function $z(y)$ in an explicit form. But
we can write the potential as a function of $y$:
\begin{eqnarray}
  V_L(z(y))&=& \eta e^{2A}\left( \eta F^2
     - \partial_y F - F\partial_{y}A \right) \nonumber \\
     &=& \frac{4\eta\lambda}{p}\cos^{-2+3/2p}(\sqrt{8p\lambda}\;y)
     \left[(8\eta+3)\sin^{2}(\sqrt{8p\lambda}\;y) -4 p\right],
     \label{VLzy}
\end{eqnarray}
where we have set $c=0$. For $0<p<3/4$, $z(y)$ is a monotonous
function, and this potential has the asymptotic behavior: $V_L(z=\pm
\infty)=0$ and $V_L(z=0)=-16\eta\lambda$. For $\eta>0$, this in fact
is a volcano type potential \cite{volcano,Davoudiasl}. This means
that the potential provides no mass gap to separate the fermion zero
mode from exited KK modes. The potential for right chiral fermions
can be obtained by the replacement $\eta \longrightarrow -\eta$ from
Eq. (\ref{VLzy}). The shape of the potentials $V_L$ and $V_R$ for
left and right chiral fermions for the case $0<p<3/4$ is plotted in
Fig. \ref{fig_VzVy_Fermion} in $y$ and $z$ coordinates.

\begin{figure}[htb]
\includegraphics[width=7.5cm,height=5.5cm]{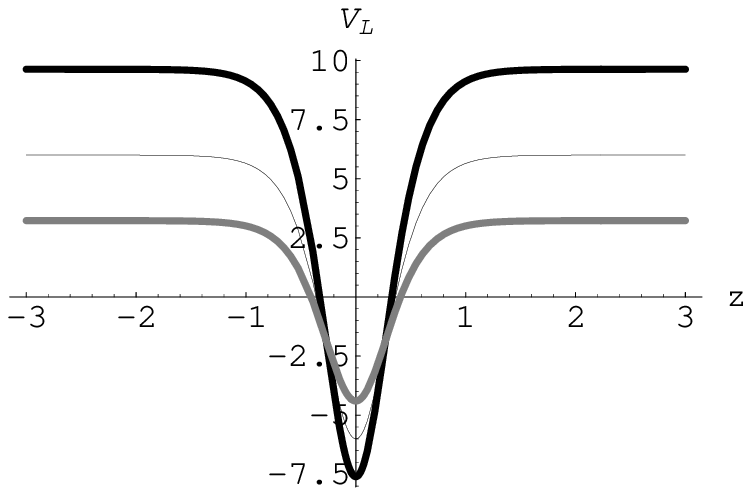}
\includegraphics[width=7.5cm,height=5.5cm]{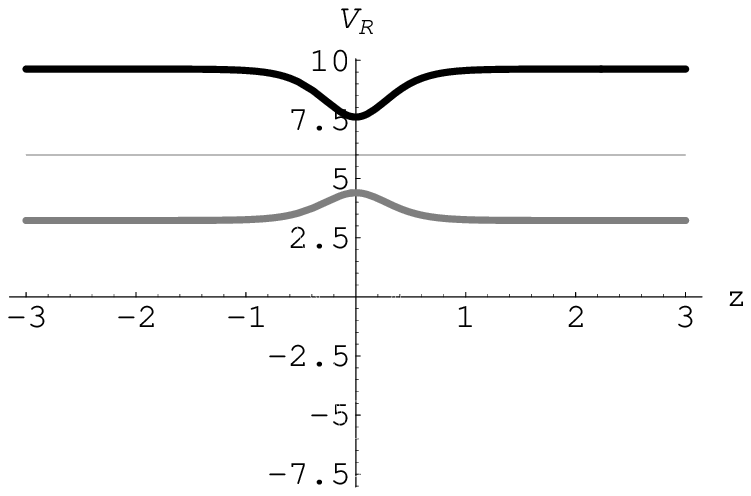}
\caption{The shape of the potentials $V_L$ and $V_R$ for left and
right chiral fermions for the case $p=3/4$. The parameters are set
to $\lambda=1$, and $\eta=3/8$ for black thin lines,
$\eta=3/8+0.1$ for black thick lines and $\eta=3/8-0.1$ for gray
thick liens.}
 \label{fig_V_fermion}
\end{figure}

Following, we mainly discuss the case $p=3/4$, for which one can
invert the coordinate transformation $dz=e^{-A(y)}dy$ and get the
explicit form of the potential for left chiral fermion
\begin{eqnarray}
 V_L(z)
 =\frac{16\eta\lambda}{3}
 \biggl[ 8\eta- (8\eta+3){\rm sech}^2(\sqrt{6\lambda}\;z) \biggl].
 ~(p=\frac{3}{4})
 \label{VeffLeftFermion}
\end{eqnarray}
For right chiral fermion, the corresponding potential can be
written out easily by replacing $\eta\rightarrow -\eta$ from above
equation
\begin{eqnarray}
  V_R(z)
  =\frac{16}{3} \eta  \lambda
 \biggl[ 8\eta- (8\eta-3){\rm sech}^2(\sqrt{6\lambda}\;z) \biggl],
 ~(p=\frac{3}{4})
  \label{VeffRightFermion}
\end{eqnarray}
and the value at $y = 0$ is given by
\begin{equation}
V_R(0) =-V_L(0) =  16 \eta\lambda.
\end{equation}
Both the two potentials have the asymptotic behavior:
$V_{L,R}(z=\pm \infty)=128\eta^2\lambda/3>0$. But for a given
coupling constant $\eta$, the values of the potentials at $z=0$
are opposite. The shape of the above two potentials is shown in
Fig. \ref{fig_V_fermion} for different values of positive $\eta$.

For positive $\eta$, only the potential for left chiral fermions has
a negative value at the location of the brane, which can trap the
left chiral fermion zero mode solved from (\ref{CoupleEq1a}) by set
$m_0=0$:
\begin{equation}
 \widetilde{\alpha}_{L0} = \left[
  \frac{\sqrt{6\lambda}\;\Gamma(\frac{8\eta}{3}+\frac{1}{2})}
       {\sqrt{\pi}\;\Gamma(\frac{8\eta}{3})}\right]^{\frac{1}{2}}
  \cosh^{-\frac{8\eta}{3}}(\sqrt{6\lambda}\;z).     ~~~~~(\eta>0)
  \label{zeroModeLeftFermion}
\end{equation}

The zero mode (\ref{zeroModeLeftFermion}) represents the lowest
energy eigenfunction of the Schr\"{o}dinger equation
(\ref{SchEqLeftFermion3}) since it has no zeros. The general bound
states for the potential (\ref{VeffLeftFermion}) can be obtained
by using the traditional recipe of transforming the stationary
Schr\"{o}dinger equation into an hypergeometric equation
\begin{eqnarray}
  \widetilde{\alpha}_{Ln}
  =\cosh^{1+\frac{8}{3}\eta} (\sqrt{6\lambda}\;z)
 \left[ d_n f_1 + g_n \sinh(\sqrt{6\lambda}\;z) f_2
  \right],~
\end{eqnarray}
where $d_n$ and $g_n$ are normalization constants and $f_1$ and
$f_2$ are the hypergeometric functions
\begin{eqnarray}
  f_1 &=& {_2}F_1 \left(a_n,b_n;\frac12;-\sinh^2(\sqrt{6\lambda}\;z)\right),
 \nonumber \\
  f_2 &=& {_2}F_1\left(a_n+\frac12,b_n+\frac12;\frac32;
                  -\sinh^2(\sqrt{6\lambda}\;z)\right)
\end{eqnarray}
with the parameters $a_n$ and $b_n$ given by
\begin{eqnarray}
 a_n &=& \frac12 \left(n+1\right), \\
  b_n &=&  \frac{8}{3}\eta -\frac12 \left(n-1\right).
\end{eqnarray}
The corresponding mass spectrum of the bound states is
\begin{eqnarray}
 m^2_n = 2\lambda(16 \eta-3n)n.~~~
 (\eta>0,~n=0,1,2,...<\frac{8}{3}\eta)~~
 \label{massSpectrumVL}
\end{eqnarray}
It is turned out that the state for $n=0$ always belongs to the
spectrum of $V_L(z)$, which is just the zero mode with $m_0=0$.
Since the ground state has the lowest mass square $m_0^2=0$, there
is no tachyonic left chiral fermion modes. Suppose $N_L-1$ is the
biggest possible value of $n$ in (\ref{massSpectrumVL}), then the
number of bound states for left chiral fermions is $N_L$. If
$0<\eta\leq 3/8$, there is just one bound state ($N_L=1$), i.e., the
zero mode, and for which we have $a_0=1/2, b_0=8\eta/3+1/2,d_0\neq
0,g_0=0$, the corresponding normalized wave function turns out to be
the zero mode given in (\ref{zeroModeLeftFermion}).
In order for the left chiral fermion potential to have at least
one bound exited state ($N_L\geq 2$), the condition $\eta>3/8$ is
needed. It is interesting to note that (\ref{VeffLeftFermion}) is
a modified transparent potential (i.e. the reflexion coefficient
is equal to zero) when $\eta=3k/8$ and $k=1,2,...$
\cite{transparentPotential}
\begin{eqnarray}
 V_L(z) =6\lambda
 \biggl[k^2-k(k+1){\rm sech}^2(\sqrt{6\lambda}\;z) \biggl],
 ~(\eta=\frac{3k}{8})
 \label{VeffLeftFermion2}
\end{eqnarray}
and there are $k$ bound states ($N_L=k$) with mass spectrum
$m^2_n=6\lambda(2k-n)n$.

\begin{figure}[htb]
\begin{center}
\includegraphics[width=7.5cm,height=5.5cm]{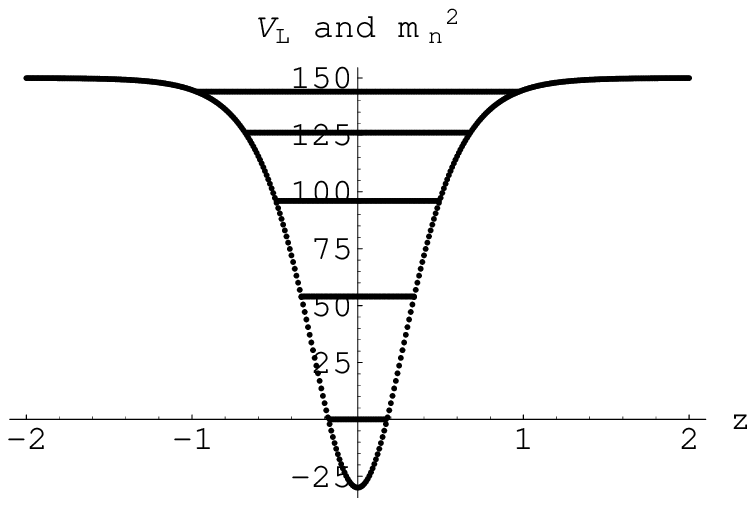}
\includegraphics[width=7.5cm,height=5.5cm]{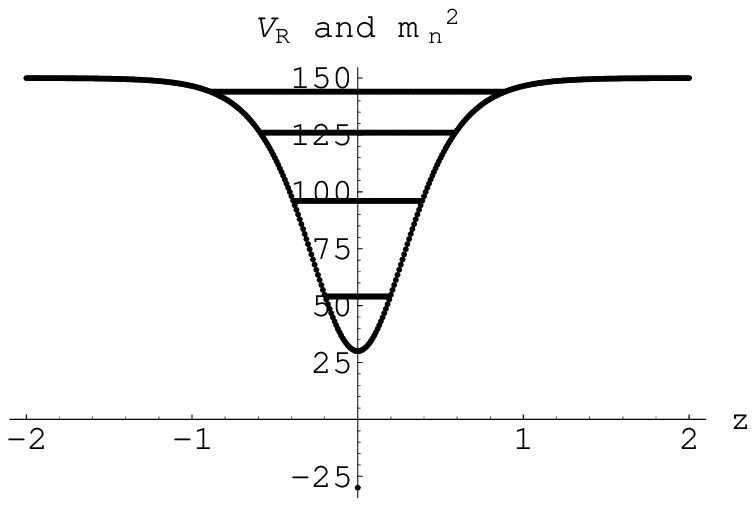}
\end{center}
\caption{The shape of the potentials $V_L$, $V_R$ and the mass
spectrum $m_n^2$ for left and right chiral fermions for the case
$p=3/4$. The parameters are set to $\eta=5\times 3/8$ and
$\lambda=1$.}
 \label{fig_Vz_Mn2_Fermion5}
\end{figure}

\begin{figure}[htb]
\begin{center}
\includegraphics[width=7.5cm,height=5.5cm]{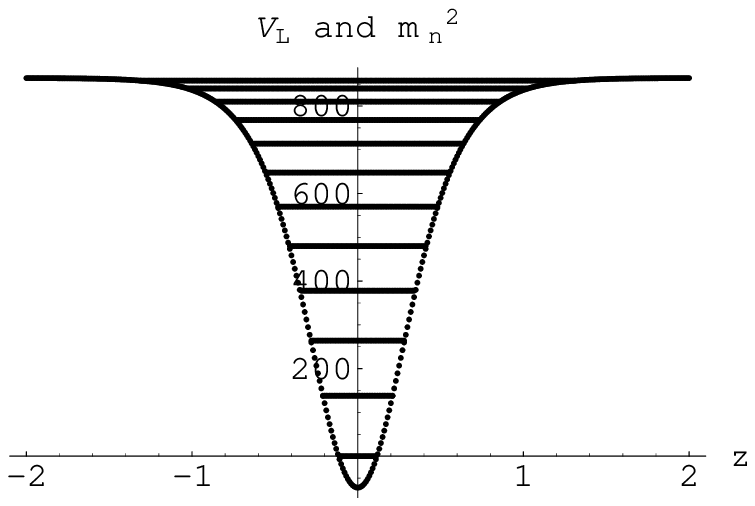}
\includegraphics[width=7.5cm,height=5.5cm]{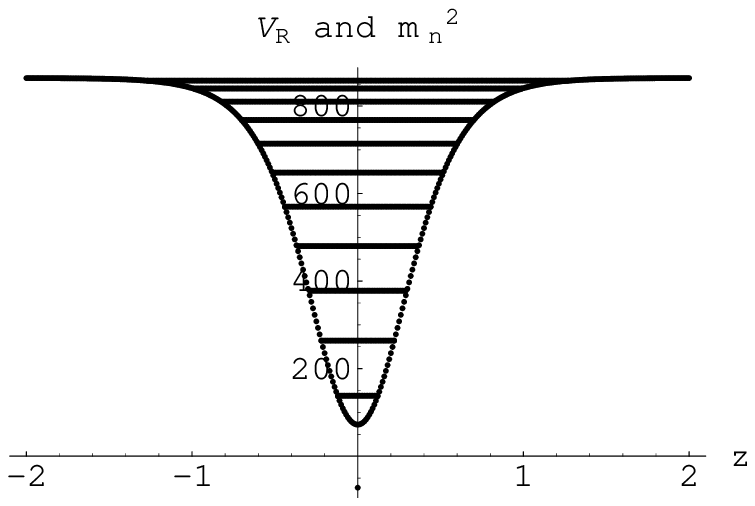}
\end{center}
\caption{The shape of the potentials $V_L$, $V_R$ and the mass
spectrum $m_n^2$ for left and right chiral fermions for the case
$p=3/4$. The parameters are set to $\eta=12\times 3/8$ and
$\lambda=1$.}
 \label{fig_Vz_Mn2_Fermion12}
\end{figure}

In the case $\eta>0$, the potential for right chiral fermions is
positive near the location of the brane, which shows that it can
not trap the right chiral zero mode. For the special value
$\eta=3/8$, the potential of right chiral fermions is a positive
constant
\begin{eqnarray}
  V_R(z)
  =6 \lambda . ~~~ (\eta=3/8)
\end{eqnarray}
For the case $0<\eta\leq3/8$, we have $V_R(0)\geq
V_R(\pm\infty)>0$, which shows that there is no any bound state
for the potential of right chiral fermions. For the case $\eta>
3/8$, $0<V_R(0)< V_R(\pm\infty)$, there is a potential barrier
which indicates that there may be some bound states, but none of
them is zero mode. The corresponding mass spectrum is
\begin{eqnarray}
\label{massSpectrumVR}
 m^2_n = 2\lambda[16 \eta-3(n+1)](n+1). \\
 (\eta>\frac{3}{8},~n=0,1,2,...<\frac{8}{3}\eta-1)\nonumber
\end{eqnarray}
The number of bound states of right chiral fermions $N_R$ is one
less than that of left chiral fermions $N_L$, i.e., $N_R=N_L-1$. If
$0<\eta\leq 3/8$, there is only one left chiral fermion bound state.
If $\eta>3/8$, there are $N_L(N_L\geq 2)$ left chiral fermion bound
states and $N_L-1$ right chiral fermion bound states. The ground
state for right chiral fermion is
\begin{equation}
 \widetilde{\alpha}_{R0} = \left[
  \frac{\sqrt{6\lambda}\;\Gamma(\frac{8\eta}{3}-\frac{1}{2})}
       {\sqrt{\pi}\;\Gamma(\frac{8\eta}{3}-1)}\right]^{\frac{1}{2}}
  \cosh^{1-\frac{8\eta}{3}}(\sqrt{6\lambda}\;z),     ~~~~(\eta>\frac{3}{8})
  \label{RightFermionGroundState}
\end{equation}
which is not any more zero mode because the mass is decided by
$m^2_0=2\lambda(16 \eta-3)>6\lambda>0$. This conclusion can be
seen clearly from the shape of $V_R$ in Fig.
(\ref{fig_V_fermion}). When $\eta=3k/8$ and $k=2,3,...$, the
potential (\ref{VeffRightFermion}) is
\begin{eqnarray}
 V_R(z) =6\lambda
 \biggl[k^2-k(k-1){\rm sech}^2(\sqrt{6\lambda}\;z) \biggl],
 ~~~~(\eta=\frac{3k}{8})
 \label{VeffRightFermion2}
\end{eqnarray}
and there are $k-1$ bound states ($N_R=k-1$) with mass spectrum
$m^2_n=6\lambda[2k-(n+1)](n+1)$. In Figs. \ref{fig_Vz_Mn2_Fermion5}
and \ref{fig_Vz_Mn2_Fermion12} we plot the shape of left and right
chiral fermions and the mass spectrum for different values of
$\eta$. For the case $\lambda=1, \eta=5\times\frac{3}{8}$, there are
5 bound states for the left fermions and 4 bound states for the
right ones and the mass spectra are
\begin{eqnarray}
 m_{Ln}^2&=&\{0, 54, 96, 126, 144\}  \cup [150,\infty), \\
 m_{Rn}^2&=&\{~~~~ 54, 96, 126, 144\} \cup [150,\infty).
\end{eqnarray}
For the case $\lambda=1, \eta=12\times\frac{3}{8}$, there are 12 and
11 bound states for the left and the right fermions respectively and
the mass spectra are
\begin{eqnarray}
 m_{Ln}^2&=&\{0, 138, 264, 378, 480, 570, 648, 714, 768, 810, 840, 858\} \cup [864,\infty), \\
 m_{Rn}^2&=&\{~~~~ 138, 264, 378, 480, 570, 648, 714, 768, 810, 840, 858\} \cup [864,\infty).
\end{eqnarray}

But for the case of negative $\eta$, things are opposite and only
the right chiral zero mode
\begin{equation}
 \widetilde{\alpha}_{R0} = \left[\sqrt{\frac{6\lambda}{\pi}}
  \frac{\Gamma(\frac{1}{2}-\frac{8\eta}{3})}
  {\Gamma(-\frac{8\eta}{3})}~\right]^{\frac{1}{2}}
  \cosh^{\frac{8\eta}{3}}(\sqrt{6\lambda}\;z)     ~~~~~~~(\eta<0)
\end{equation}
can be trapped on the brane. For arbitrary $\eta\neq 0$, the two
potentials suggest that there exist mass gap (at least for one of
them) and a continuous spectrum of KK modes with positive $m^2>0$
(for both of them), which are same as the cases of scalars and
vectors obtained in the section. It is worth noting that, in the
case of no coupling ($\eta=0$), both the two potentials for left
and right chiral fermions are vanish, and hence there are no any
localized fermion KK modes including zero modes.

Localization of fermions in general spacetimes has been studied for
example in \cite{RandjbarPLB2000}. In Ref. \cite{MelfoPRD2006},
Melfo {\em et al } showed that only one massless chiral mode is
localized in double walls and branes interpolating between different
$AdS_5$ spacetimes whenever the wall thickness is keep finite, while
chiral fermionic modes cannot be localized in $dS_4$ walls embedded
in a $M_5$ spacetime. Localizing the fermionic degrees of freedom on
branes or defects requires us to introduce other interactions but
gravity. Recently, Parameswaran {\em et al} studied fluctuations
about axisymmetric warped brane solutions in 6-dimensional minimal
gauged supergravity and proved that, not only gravity, but Standard
Model fields could be described by an effective 4-Dimensional theory
\cite{Parameswaran0608074}. Moreover, there are some other
backgrounds such as gauge field \cite{LiuJHEP2007}, supergravity
\cite{Mario} and vortex background
\cite{LiuNPB2007,LiuVortexFermion} could be considered. The
topological vortex coupled to fermions may result in chiral fermion
zero modes \cite{JackiwRossiNPB1981}. More recently, Volkas {\em et
al } had extensively analyzed localization mechanisms on a domain
wall. In particular, in Ref. \cite{Volkas0705.1584}, they proposed a
well-defined model for localizing the SM, or something close to it,
on a domain wall brane.

\section{Discussions}

In this paper, we have investigated the possibility of localizing
various matter fields on a Weyl thick brane, which also localize the
graviton, from the viewpoint of field theory. We first give a brief
review of the type of thick smooth brane configuration in a pure
geometric Weyl integrable 5-dimensional space time. Then, we check
localization of various bulk matter fields on the pure geometrical
thick brane and obtain the KK spectrums for the mass-independent
potential of these matter fields. When $0<p<3/4$, the one
dimensional Schr\"{o}dinger potentials for scalars, vectors and
fermions are similar to the one for gravity obtained in Ref.
\cite{ThickBrane4}. They have a finite negative well at the location
of the brane and a finite positive barrier at each side which
vanishes asymptotically. It is shown that there is only one single
bound state (zero mode) which is just the lowest energy
eigenfunction of the Schr\"{o}dinger equation for the three kinds of
fields. Since all values of $m^2>0$ are allowed, there also exist a
continuum gapless spectrum of KK states with $m^2>0$, which turn
asymptotically into continuum plane wave as $|z|\rightarrow \infty$
\cite{Lykken,dewolfe,ThickBrane2,ThickBrane3}. All of these zero
modes including the one for spin 1 vectors are normalized and bound,
so all these matter fields are localized on the brane.

When $p=3/4$, the potentials are the modified P\"{o}schl-Teller
potentials. They are also similar to the case of gravity and have a
finite negative well at the location of the brane and a finite
positive barrier at each side which doesn't vanishes. These
potentials suggest that there exist mass gap and a series of
continuous spectrum starting at positive $m^2$. The discrete KK
modes are bound states while the continuous ones are not. For
scalars, there are two bound KK modes, which is just same as the
case of gravity. For spin one vectors, there is only one bound
state, which is the zero mode. The total number of bound states for
spin half fermions is determined by the coupling constant $\eta$.
For positive coupling constant, the number of bound states of right
chiral fermions is one less than that of left chiral fermions. If
$0<\eta\leq 3/8$, there is only one left chiral fermion bound state
which is just the left chiral fermion zero mode. If $\eta>3/8$,
there are $N_L(N_L\geq 2)$ left chiral fermion bound states
(including zero mode and massive KK modes) and $N_L-1$ right chiral
fermion bound states (only including massive KK modes). For negative
coupling constant, we will get similar results but need to
interchange left and right, e.g., there is the localized right
chiral fermion zero mode but not the localized left one. In the case
of no coupling ($\eta=0$), there are no any localized fermion KK
modes including zero modes for both left and right chiral fermions.
Hence, for left or right chiral fermions localization, there must be
some kind of coupling. These situations can be compared with the
case of the domain wall in the RS framework \cite{BajcPLB2000},
where for localization of spin 1/2 field additional localization
method by Jackiw and Rebbi \cite{JackiwPRD1976} was introduced.

\section*{Acknowledgement}
It is a pleasure to thank the authors of Ref. \cite{ThickBrane2} for
their very helpful and interesting discussion. This work was
supported by the National Natural Science Foundation of the People's
Republic of China (No. 502-041016, No. 10475034 and No. 10705013)
and the Fundamental Research Fund for Physics and Mathematics of
Lanzhou University (No. Lzu07002).

\end{document}